# Multiwavelength electron diffraction as a tool for identifying stacking sequences in 2D materials

Pascal Puech[1*], Iann Gerber[2], Fabrice Piazza[3], Marc Monthioux[1*]

[1]Centre d'Elaboration des Matériaux et d'Etudes Structurales (CEMES), UPR-8011 CNRS, Université de Toulouse, France.

[2]Laboratoire de Physico-Chimie des Nano-Objets (LPCNO), UMR 5215 CNRS, INSA, Université de Toulouse, France.

[3]Laboratorio de Nanociencia, Pontificia Universidad Católica Madre y Maestra, Santiago de Los Caballeros, Dominican Republic.

*Corresponding authors (pascal.puech@cemes.fr; marc.monthioux@cemes.fr)

**Two-dimensional (2D) materials are among the most studied ones nowadays, because of their unique properties. These materials are made of, single- or few atom-thick layers assembled by van der Waals forces, hence allowing a variety of stacking sequences possibly resulting in a variety of crystallographic structures as soon as the sequences are periodic. Taking the example of few layer graphene (FLG), it is of an utmost importance to identify both the number of layers and the stacking sequence, because of the driving role these parameters have on the properties. For this purpose, analysing the spot intensities of electron diffraction patterns (DPs) is commonly used[1-9] along with attempts to vary the number of layers[1,7] and the specimen tilt angle[2-4]. However, the number of sequences able to be discriminated this way remains few, because of the similarities between the DPs. Also, the possibility of the occurrence of C layers in addition to A and/or B layers in FLG has been rarely considered[1,6,8]. To overcome this limitation, we propose here a new methodology based on multi-wavelength electron diffraction which is able to discriminate between stacking sequences up to 6 layers (potentially more) involving A, B, and C layers. We also propose an innovative method to calculate the spot intensities in an easier and faster way than the standard ones. Additionally, we show that the method is valid for transition metal dichalcogenides, taking the example of $MoS_2$.**

For 2D crystallites, the intensity shown on the diffraction pattern (DP) for a given spot not only depends on the number of atoms involved in the scattering process, but also on where the Ewald sphere surface intersects the reciprocal rod (**Fig.1**).

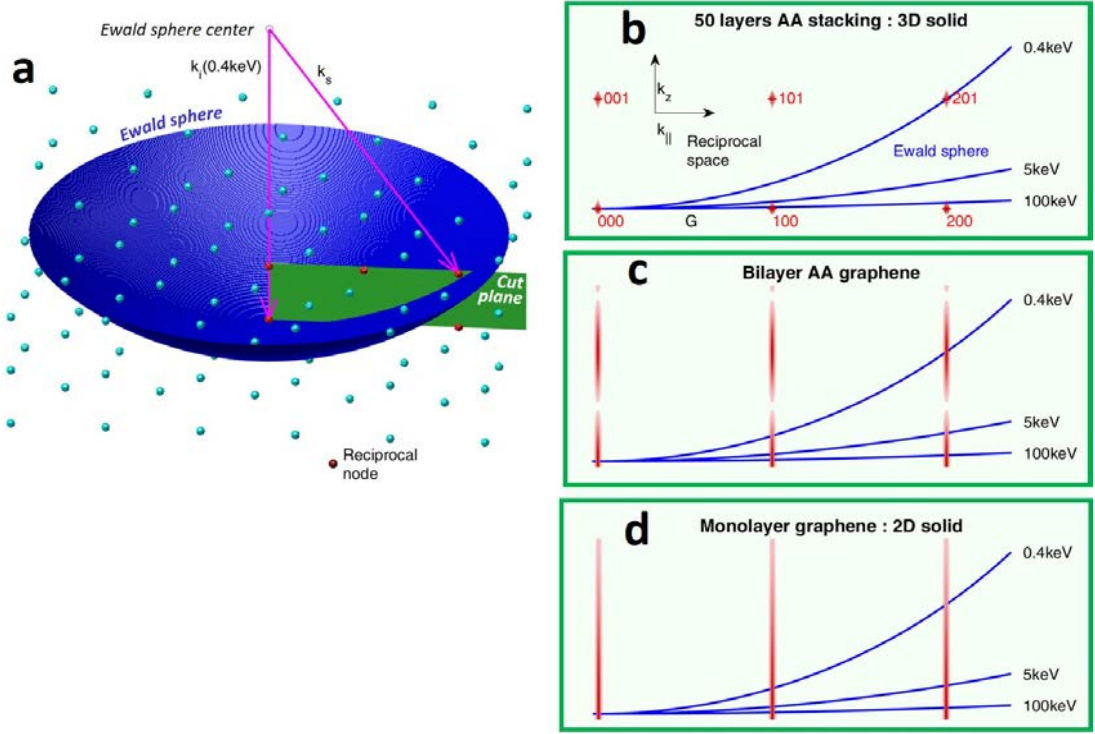

**Figure 1. The electron wave/crystallite interaction in the reciprocal space**. **(a)** Sketch of the reciprocal image of the electron wave (the Ewald sphere) interacting with the reciprocal image of a 50 AA-stacked layer graphene. A long wavelength ($\lambda = 0.613$ Å, 0.4 keV) was chosen so that to enhance the sphere curvature. **(b)** to **(d)** represent the green plane in (a) while considering an electron energy of 0.4, 5, and 100 keV, respectively. **(b)** When the crystallite is large and 3D (here the same graphene as in (a)), reciprocal nodes are nearly dimensionless, and one spot intensity is proportional to the number of atoms involved in the scattering process for the scattering direction considered. **(d)** When a crystallite is thinned down to a single layer (here a monolayer graphene), resulting in a 2D crystallite, reciprocal nodes elongate in the thinned direction, and the potentially-scattered intensity distributes along each rod in a Gaussian way. **(c)** For a limited number of layers coherently piled up (here a 2-layer graphene in AA stacking), reciprocal rods become modulated to account for the periodicities in the z direction. The higher the number of layers, the less elongated the reciprocal nodes in the z direction. Sketches are at scale. The AA structure was chosen so that both *h00* and *h01* nodes be contained in the green plane. Graphenes for the sketches are considered free of any corrugation[2] or deformation. Further comments are provided in Supplementary Information.

For an electron energy of 100 keV, the Ewald sphere is large enough for its surface to be approximated as a plane (**Figs.1b-d**). For a given electron energy, exploring the reciprocal space is only possible by tilting the sample[2,3]. However, the lower the electron energy, the shorter the Ewald sphere radius, inducing the Ewald sphere surface to intersect the various reciprocal rods at different heights (**Figs.1b-d**), thereby generating diffraction patterns exhibiting the same spots but with variable intensities, or even with some spots missing. Our study reveals that the spot intensities may vary dramatically with the electron energy, which has never been considered in the literature despite a range as large as from 42 eV[10] to 800 keV[1] has been used. Nowadays, a majority of papers dealing with graphene are using energies in the 60-100 keV range in order to avoid or limit irradiation damaging, since ~80 keV was found to be enough for promoting atom knock-on events[11]. But the fact is that the influence of the electron energy on spot intensities in DPs is neglected, and mentioning the acceleration voltage value may even be omitted in article main texts[3,4,12].

In order to investigate the influence of the electron energy on the scattered intensities for 2D materials, we have simulated a variety of DPs first considering graphene systems made of a variety of number of layers $N$ (from 1 to 6) under various stacking configurations, such as A, AA, AB, ABA, ABC, ABAB, *etc*, while using a decreasing electron energy ranging from 100 to 5 keV. The relative positions of layers A, B, and C are reminded in **Supplementary Information** (SI) and **Fig.S1b**. ABA corresponds to the stacking sequence in the unit cell of the hexagonal structure of graphite, whereas ABCA corresponds to the stacking sequence in the unit cell of the rhombohedral structure of graphite.

DPs were obtained thanks to innovative calculation principles as follows (also see Section 2 in **SI**):

$$S_{electron}(\vec{q}, \theta_{\vec{q}}) = B \frac{1}{N} \left| \sum_{el_i=1}^{N_{el}\ in\ the\ primitive\ cell} f_{el_i}(q) \sum_{i \in el_i} \exp(i[2\pi(hx_i + ky_i) + q_z z_i c]) \right|^2$$

where $x_i$, $y_i$ and $z_i$ are the relative coordinates in the cell. Taking graphene as an example, it corresponds to 2 atoms in each plane when all planes are stacked in the $z$ direction, where $h$ and $l$ are the Laue indices. $q_z$ is the vector in the reciprocal space involved in the diffraction (see **SI**). With simple algebra, we obtain:

$$q_z = \frac{2\pi}{\lambda} - \sqrt{\left(\frac{2\pi}{\lambda}\right)^2 - G^2}$$

Where $G$ is a vector of the reciprocal space in the $(k_x k_y)$ plane.

The reliability and relevance of the calculation principles described above were validated by comparing with the results obtained by a regular DP calculation software such as XaNSoNS (corrected with the contribution of the electron clouds, as this software is initially devoted to X-ray and neutron scattering[13]) whose principles are based on considering all atoms and all directions in the reciprocal space, which makes calculations heavier (although massive parallelisation can help) (see **SI**). Both the methodology we propose and that of XaNSoNS are based on the real part of the atomic structure factor. Ideally, the latter should be corrected from the anomalous scattering, leading to the additional contribution of the imaginary part (corresponding to the inelastic scattering). But the atomic structure factor for single-element materials such as graphene can be factorised, and thus, only a very subordinate intensity variation for each ring is expected. Our original calculation allows exploring the effect of varying the electron energy on the intensity associated to nodes in the reciprocal space in selected directions.

As a matter of example, **Fig.2** provides calculated DPs obtained at 5 keV electron energy for three selected configurations along with the corresponding experimental DPs obtained from two-layer graphene (2LG) materials, also at 5 keV. The spot intensity features for single graphene (**Fig.2a**) appears similar to what is reported in the literature for DP at 60-200 keV[1,2,14,15] but (1) the intensity difference between Ring1 and Ring2 in 2LG-AA (**Fig.2b**), and (2) the evidence of the three-fold symmetry in the spot intensity distribution on the first ring for 2LG-AB (**Fig.2c**), are both unprecedented observations.

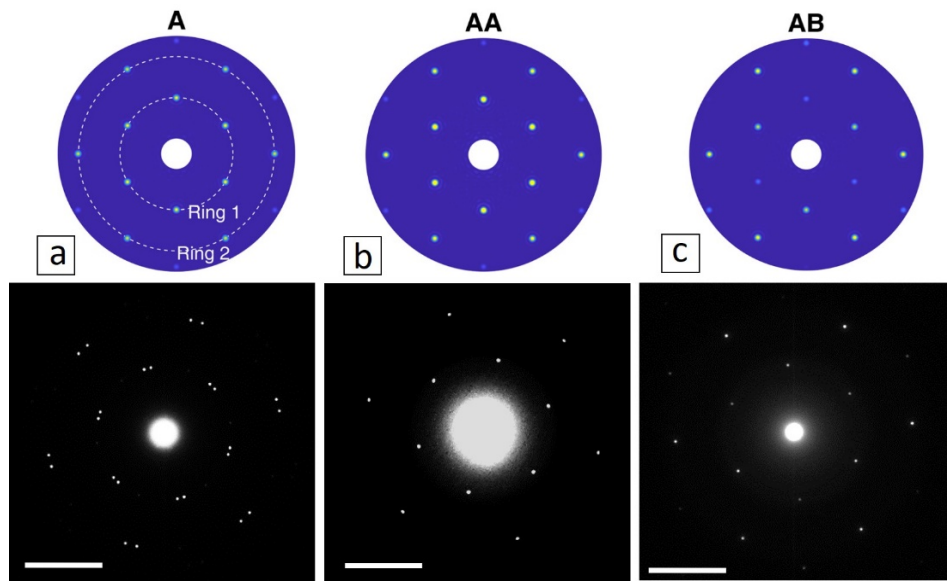

**Figure 2. Electron diffraction patterns of 2LG at low voltage.** Examples, for a 5 keV electron energy, of calculated (above, using XaNSoNS, corrected by considering the electron cloud) and experimental (below, on 2LG materials) electron DPs. **(a)** single graphene, where all spot intensities on Ring1 are equal and also equal to that on Ring2; the experimental DP actually exhibits two monolayers incoherently stacked, since twisted with a 6° angle. **(b)** 2LG-AA, where all spot intensities on Ring1 are equal and slightly higher than that on Ring2. **(c)** 2LG-AB, where the spot intensities on Ring1 are distributed according to a three-fold symmetry, i.e. with an alternance of strong and weak spots,

whereas all spot intensities on Ring1 are lower than that on Ring2. Actual intensity values are provided in **Table1**. Scale bars are 5 nm$^{-1}$.

Examples of how the intensity features in the calculated DPs vary for the two rings, Ring1 and Ring2, for a selection of different stacking sequences when varying the electron energy continuously from 1 to 100 keV are provided in **Fig.3** (others are provided in **SI**). Major discrepancies are observed, even for configurations which were supposed to be structurally close, such as A and AA, or AB and ABA. In particular, AB exhibits a three-fold symmetry of the spot intensity distribution, while ABA does not. Also, the ABC sequence barely exhibits Ring1 at high electron energy.

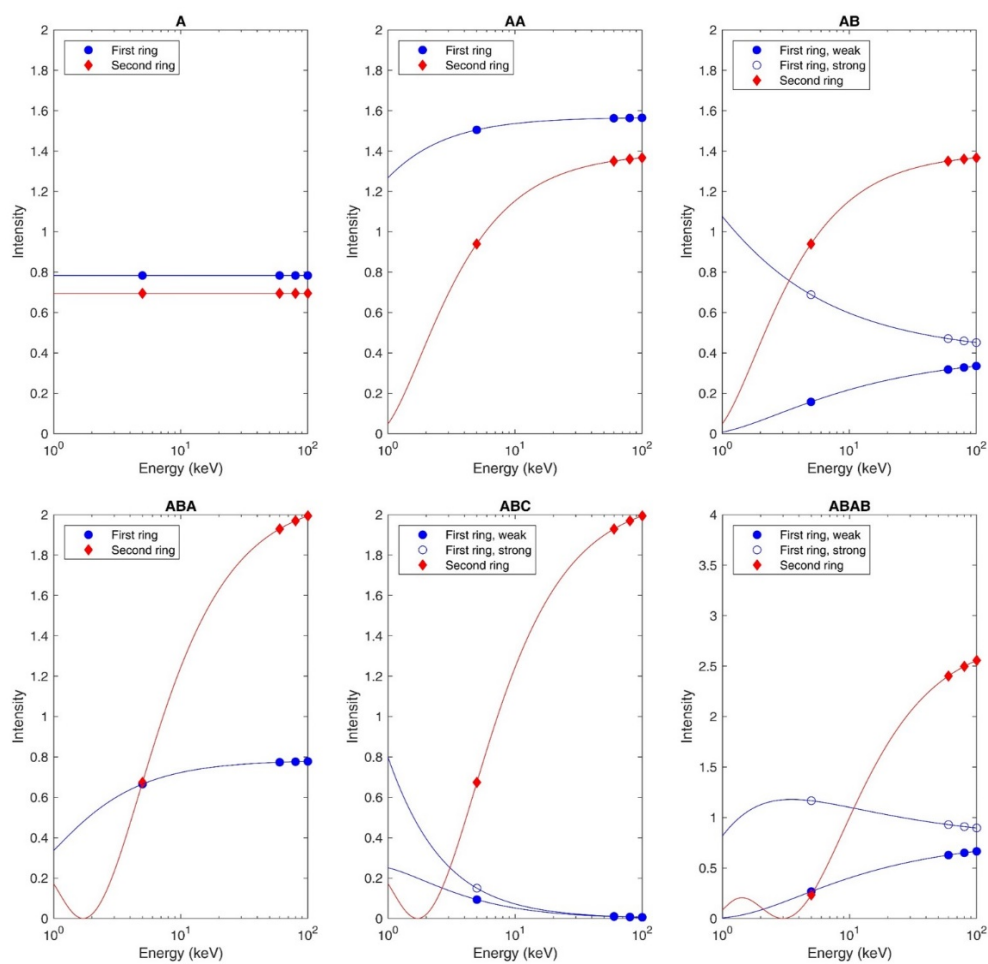

**Figure 3. Variation of the diffracted spot intensities with electron energy and stacking sequence type in graphene**. Intensity variations for a single spot from each ring, Ring1 (blue line) and Ring2 (red line), in calculated DPs with respect to the continuous variation of the electron energy (from 1 to 100 keV) for a selection of stacking configuration in graphene. When a three-fold symmetry of the intensity distribution appears on Ring1, two intensity values are provided corresponding to that of one weak spot (solid circle) and one strong spot (open circle), respectively. Electron energies corresponding to 5, 60, 80, and 100 kV TEM acceleration voltage are marked, for easier recognition.

**Table 1** reports all the calculated intensity values for all the eleven stacking sequences studied for both 5 and 100 keV for comparison. It appears that the three-fold symmetry of the intensity distribution on Ring1 at 5 keV is typical of graphenes in Bernal stacking sequence with an even number of layers. On the other hand, as soon as a C layer is involved, the three-fold symmetry appears systematically regardless of the number of layers, although sometimes with an intensity difference too faint for being noticeable. Interestingly, a given stacking sequence which does not exhibit any singularity at high electron energy value, making difficult its identification, can exhibit one at low electron energy value, or the contrary. For instance, the three-fold symmetry of the intensity distribution on Ring1 barely shows up at 100 keV ($I_{Ring1(strong)}/I_{Ring1(weak)}$ = 1.8 at maximum) whereas it is more obvious at 5 keV ($I_{Ring1(strong)}/I_{Ring1(weak)}$ goes up to 10 or even infinite). Therefore, identifying each of the stacking sequences, which was mostly impossible so far when only using the standard 100 keV electron energy, now appears to be possible by combining DPs at low and high electron energies when needed. It therefore comes out that a Table such as **Table 1** can play the role of an identification chart.

| Stacking sequence | 5 keV | | | | | 100 keV | | | | |
|---|---|---|---|---|---|---|---|---|---|---|
| | Ring1, weak | Ring1, strong | Ring1$_{(S)}$ / Ring1$_{(W)}$ | Ring2 | Ring2 / Ring1$_{(S)}$ | Ring1, weak | Ring1, strong | Ring1$_{(S)}$ / Ring1$_{(W)}$ | Ring2 | Ring2 / Ring1$_{(S)}$ |
| A | 0.784 | | 1 | 0.695 | 0,9 | 0.784 | | 1 | 0.695 | 0.9 |
| AA | 1.505 | | 1 | 0.940 | 0.6 | 1.564 | | 1 | 1.367 | 0.9 |
| AB | 0.157 | 0.689 | 4.4 | 0.940 | 1.4 | 0.335 | 0.451 | 1.3 | 1.367 | 3 |
| ABA | 0.665 | | 1 | 0.674 | 1 | 0.778 | | 1 | 1.994 | 2.6 |
| ABAB | 0.267 | 1.166 | 4.4 | 0.234 | 0.2 | 0.665 | 0.896 | 1.3 | 2.556 | 2.9 |
| ABABA | 0.736 | | 1 | 0.006 | ~0 | 1.078 | | 1 | 3.036 | 2.8 |
| ABABAB | 0.299 | 1.309 | 4.4 | 0.079 | ~0 | 0.986 | 1.328 | 1.3 | 3.421 | 2.6 |
| ABC | 0.09 | 0.150 | 1.7 | 0.674 | 4.5 | 0.005 | 0.006 | ~1 | 1.994 | 332 |
| ABCA | 0.020 | 0.201 | 10 | 0.234 | 1.2 | 0.162 | 0.220 | 1.4 | 2.556 | ~12 |
| ABCAB | 0.000 | 0.220 | ∞ | 0.006 | ~0 | 0.109 | 0.200 | 1.8 | 3.036 | 15 |
| ABCABC | 0.127 | 0.203 | 1.6 | 0.079 | 0.4 | 0.011 | 0.012 | 1 | 3.421 | 285 |

**Table 1. Identification chart for a series of possible stacking sequences in graphene up to 6 layers.** Intensities reported are that of a single spot for each spot series (Ring1$_{(weak)}$, Ring1$_{(strong)}$, Ring2). It is considered that an intensity ratio in the range ~0.7-1.5 cannot provide an intensity difference high enough for being visible to eye, and for not to be misled by intensity variations due to local deformation[3]. Values in green-shaded boxes are the parameters whose combination allows the various stacking sequences to be univocally discriminated. Detailed comments are provided in Supplementary Information.

From **Table 1**, it appears that DPs at 100 keV are never enough to determine stacking sequences with no ambiguity. Identifying various stacking sequences of graphene with a layer number up to seven was obtained once at 100 keV[6] but it was only possible because they all were pertaining to the same flake and that the neighbouring stacking sequences were needed to be taken as references. On the other hand, taking DPs at 5 keV can be enough to identify some of the stacking sequences (AA, ABABAB, ABCAB) in an absolute way. Of course, Determining the number of layers in an independent manner might change this statement. Importantly, the presence of C layers can be recognised, which has been scarcely considered so far in the literature when investigating few-layer graphenes (FLG)[1,6,8], whereas such an occurrence is possible for FLGs obtained from graphite exfoliation.

Similar statements were found for $MoS_2$, indicating that the methodology should also be valid for other 2D-materials such as TMDs, which are also materials resulting from the assembly of layers by van der Waals forces. Consequently, as in graphene, several types of stacking sequences may occur. The projected hexagonal, tetragonal, or rhombohedral structures of $MoS_2$ along $z$ direction make DPs exhibit similar features as for graphene such as two main rings wearing 6 spots each. Then, the same methodology as we proposed for graphene can be used.

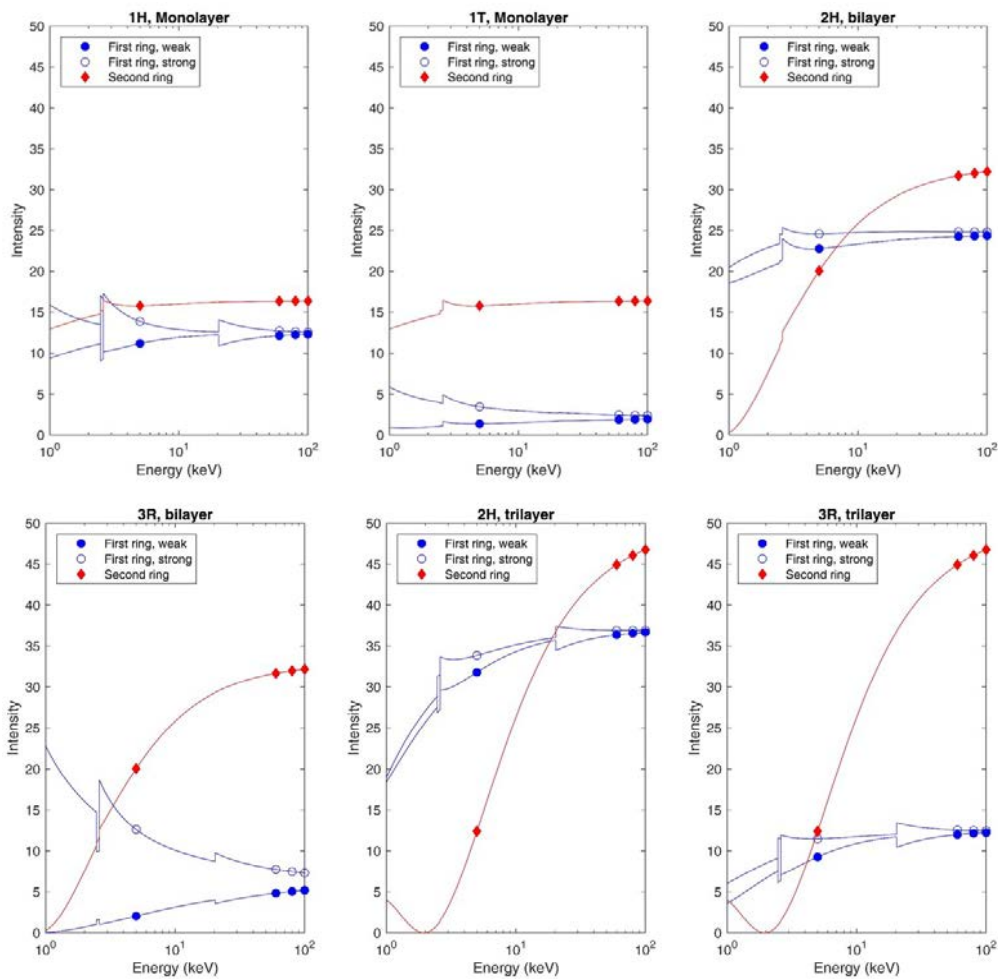

**Fig. 4. Variation of the diffracted spot intensities with electron energy and stacking sequence type in MoS$_2$.** Intensity variations for a single spot from each ring, Ring1 (blue line) and Ring2 (red line), in calculated DPs with respect to the continuous variation of the electron energy (from 1 to 100 keV) for a selection of stacking configuration in MoS$_2$. Because of the systematic three-fold symmetry of the intensity distribution on Ring1, two intensity values are provided corresponding to that of one weak spot (solid circle) and one strong spot (open circle), respectively. Electron energies corresponding to 5, 60, 80, and 100 kV TEM acceleration voltage are marked, for easier recognition.

Because each layer includes 3 sub-layers (1 Mo layer between 2 S layers) within which the relative spatial display of the atoms may vary, there are more structure variations than in graphene, and the overall atomic structure factor is more complicate than for pure carbon material. Due to the higher atomic numbers of the atoms involved, the energy of the incident wave in the 1-100 keV range can overlap some of the absorption edges of Mo and S (see **Fig.S4** in SI), leading to a larger imaginary part of the atomic structure factor. The corrections related to the inelastic scattering can be easily incorporated with our methodology and we have included them as explained in the SI. Because of the existence of the two Mo and S sublattices, the lattice symmetry is reduced from 6-fold to 3-fold which can be used to obtain the crystallite orientation.[16,17,18]

Because of the variety of cases, we will not go here to the same detailed analysis as for the graphene system above, but looking at **Fig.4**, which provides few examples of similar plots as **Fig.3**, clearly indicates that similar large intensity variations occurs when varying the electron energy over the 1-100 keV range, allowing phase identification when DP features at a single energy does not allow it. For instance, **2H bilayer** does not distinguish from **2H trilayer** at 60-100 keV, but considering DPs at 5 keV makes it possible. Same for **3R trilayer** and **3R bilayer**, thanks to the enhancement of a three-fold symmetry of the intensity distribution on Ring1 at 5 keV. The enhancement of the three-fold symmetry of the spot intensity distribution for MoS$_2$ in DPs by using low energy electrons was already observed by a surface diffraction method (µLEED) using electron energies lower than 100 eV[19]. Two accidents are seen to perturbate the intensity versus energy curves at ~2.5 and 20 keV. They correspond to the energy values of the K and L adsorption edges for both S and Mo (see Fig.S4 in SI). This (1) confirms that the imaginary part of the atomic structure factor should definitely be considered for 2D materials containing heavy atoms such as TMDs, and (2) indicates that, depending on the type of TMD to be studied, the electron energies to be used for obtaining the DPs should be suitably selected to avoid overlapping the heavy atom adsorption edges.

The consequences of the findings above are important with respect to *(i)* the current literature, *(ii)* the electron diffraction methodology, *(iii)* 2D-material identification. They mean that applying identification recipes proposed in the literature for a given electron energy to DPs obtained with another electron energy could drive to wrong results. This statement, however, does not apply for electron energies ranging from 60 to 100 keV, which are values commonly used nowadays and which provide similar results, but applies when comparing the latter energy range with energy below 10 keV. On the other hand, using high energies (e.g., above 60 keV) generally does not allow discriminating between the various possible stacking sequences. Unless intensity differences between diffracted spots are fairly large for a given energy, attempts to rely on an intensity measurement equipment for identifying the actual structure of 2D materials are risky because the thin film nature of the materials makes them likely to easily deform. Because of the elongated nature of the reciprocal nodes (**Fig.1c-d**), such a film deformation can make the elongated reciprocal rods to intersect the Ewald sphere at intensity values which are different from the nominal ones, inducing misleading intensity ratios[20], hence possibly leading to wrong identification. The risk is therefore much lowered by combining DPs obtained at both low and high electron energies. Moreover, using multiple electron energy diffraction appeared to be able to discriminate between various stacking sequences which were not able to be discriminated so far, when using single-energy diffraction. Such a possibility is important when it is needed to identify specific graphene materials, *i.e.* with the right stacking sequence, for example for allowing their subsequent transformation upon appropriate treatments to obtain new materials such as diamane[21].

The most studied 2D-materials nowadays, namely graphene and TMDs, appear to be sensitive to electron irradiation damages, leading material scientists to operate TEM at lower and lower electron energies, which the progresses in optoelectronics allow, thanks to aberration corrector systems. In spite of this, the incidence of the electron energy value on the characteristics of the DPs is not considered indeed. Our findings not only point out the need to check on the incidence of the incoming electron energy on the DP features, but also go farer by proposing multi-wavelength electron diffraction as a new and efficient way to discriminate between 2D-materials of the same family regarding the number of layers involved and their related stacking sequences.

## Methods

**2LG material.** As-received suspended bi-layer (2LG) graphene films grown by chemical vapor deposition and deposited onto 3 mm-diameter gold Quantifoil TEM grids were supplied from *Graphenea* (https://www.graphenea.com). The films are polycrystalline and were obtained from the successive transfer of two individual single-layer graphene. The resulting 2LG stacking sequence type therefore varies randomly across the film, with crystallite sizes not larger than 20 µm.

**Electron diffractions.** They were carried-out with a low voltage (5 kV) benchtop transmission electron microscope from *America Delong*. Electron source is a Schottky-type field-emission gun. Electron diffraction patterns (DPs) were obtained in selected area mode from 100 nm-large areas.

**Calculations.** The principles and basic equations are provided in the Supplementary Information.

## Acknowledgements

The authors thank R. Ruoff for stimulating discussions. This work is founded by the EUR grant NanoX n° ANR-17-EURE-0009 in the framework of the "*Programme des Investissements d'Avenir*" and by the Ministry of Higher Education, Science and Technology of the Dominican Republic (2016-2017 and 2018 FONDOCyT programs).

## Authors contributions

P.P. is at the origin of the methodology principle and performed the calculations. I.G. contributed to the calculations and supplied information on MoS$_2$. F.P. provided the 2LG materials and carried-out the low-voltage electron diffraction work. M.M. wrote the paper. All the authors contributed to discussing the results and the manuscript versions.

## Additional information

Supplementary Information are provided, reminding the basics on the possible stacking sequences in graphene, and giving details on the methodology proposed to calculate spot intensities in DPs.

## Competing financial interests

The authors declare no competing financial interests.

# Multiwavelength electron diffraction as a tool for identifying stacking sequences in 2D materials

Pascal Puech[1*], Iann Gerber[2], Fabrice Piazza[3], Marc Monthioux[1*]

[1]Centre d'Elaboration des Matériaux et d'Etudes Structurales (CEMES), UPR-8011 CNRS, Université de Toulouse, France.

[2]Laboratoire de Physico-Chimie des Nano-Objets (LPCNO), UMR 5215 CNRS, INSA, Université de Toulouse, France.

[3]Laboratorio de Nanociencia, Pontificia Universidad Católica Madre y Maestra, Santiago de Los Caballeros, Dominican Republic.

**I – Reminders on the wave/crystallite interaction in the reciprocal space**

When a crystallite interacts with a single-wavelength and a single-phase incident wave, the electrons (or photons) of the wave are diffracted, i.e., scattered coherently in specific directions of space. This interaction in the real space has a counterpart in the reciprocal space where the diffraction pattern takes place, and the latter results from the intersection of the reciprocal image of the incident wave, so-called the Ewald sphere (whose radius $ki$ is $2\pi/\lambda$, sketched by the dark blue surface in **Fig.1a** in the main article), with the reciprocal lattice of the crystallite, consisting of reciprocal nodes – sketched as light blue or red balls in **Fig.1a** in the main article. Each node represents one family of atom planes in the crystallite in real space (red-coloured nodes are those which intersect the green cut-plane in **Fig.1a**). The reciprocal nodes carry the intensity potentially scattered in the direction they represent (i.e., the direction of a family of atom planes, added with the scattering angle). Every reciprocal node whose position in the reciprocal space makes it intersect the Ewald sphere surface generates a diffracted beam, resulting in a spot in the diffraction pattern.

**II – Stacking sequences**

There are only three possibilities for three graphene layers to stack coherently, AAA, ABA, and ABC, respectively. It is reminded that, to obtain the B position starting from the A position, a shift of the layer over a distance equal to a C-C bond length in the direction of one of the in-plane C-C bonds is needed (**Fig.S1b**). Likewise, starting from the B position, the C position is obtained by keeping shifting the layer in the same direction of an additional C-C bond length distance (**Fig.S1c**). ABA corresponds to the stacking sequence in the unit cell of the hexagonal structure of graphite, whereas ABCA corresponds to the stacking sequence in the unit cell of the rhombohedral structure of graphite. The latter has however never been found as isolated crystals, but is frequently present in graphite as stacking faults instead[1].

Top views of each stacking type are sketched in **Fig.S1** below:

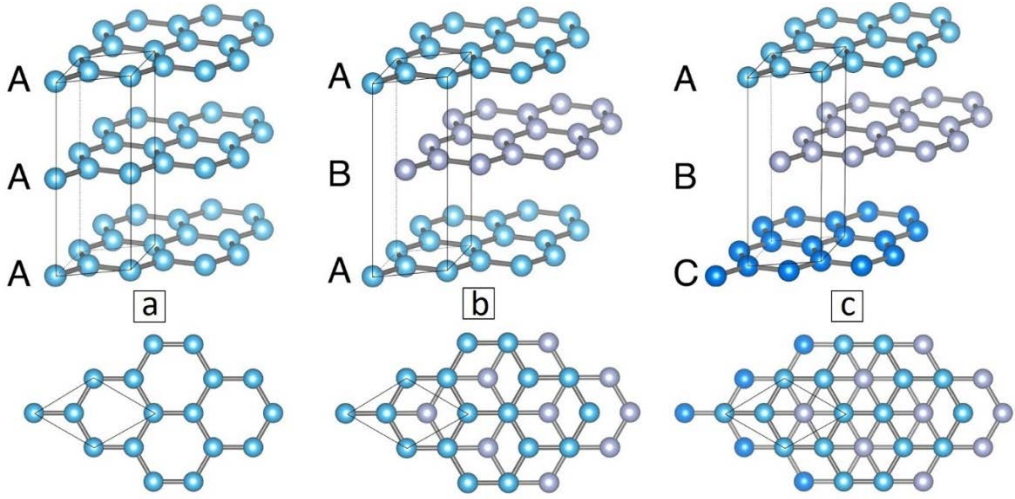

**Fig.S1.** top view of the three possible coherent stacking sequences for a graphene with N = 3 layers. **(a)** AAA stacking (bottom image also valid for any stacking of A-type layers, whatever N); **(b)** ABA stacking (bottom image also valid for any stacking sequence of A- and B-type layers, whatever N); **(c)** ABC stacking (bottom image also valid for any stacking sequence of A-, B-, and C-type layers, whatever N).

### III - Principle of calculation

The 2D pattern can be obtained through:

$$S_{electron}(\vec{q},\theta_{\vec{q}}) = A\frac{1}{N}\left|\sum_{el_i=1}^{N_{el}\ in\ the\ solid} f_{el_i}(q) \sum_{i \in el_i} \exp(i\vec{q}.\vec{R_i})\right|^2$$

In our case, with 2D systems, we introduce $\vec{R_i} = n_{a1}\vec{a_1} + n_{a2}\vec{a_2}+\vec{r_i}$ where $\vec{a_1}$ and $\vec{a_2}$ are the lattice vector in the direct space and $\vec{r_i}$ is the position in the cell primitive in the *(x,y)* plane and having a ength along z corresponding to the number of layers. Therefore, we obtain:

$$S_{electron}(\vec{q},\theta_{\vec{q}}) = A\frac{1}{N}\left|\sum_{n_{a1}=0}^{\infty}\sum_{n_{a1}=0}^{\infty} \exp(i\vec{q}.(n_{a1}\vec{a_1}+n_{a2}\vec{a_2})) \times \left\{\sum_{el_i=1}^{N_{el}\ in\ the\ cell} f_{el_i}(q) \sum_{i \in el_i} \exp(i\vec{q}.\vec{r_i})\right\}\right|^2$$

*Supplementary information*

Here, the first double sum gives:

$$\left|\sum_{n_{a1}=0}^{n_\infty}\sum_{n_{a1}=0}^{n_\infty} \exp(i\vec{q}\cdot(n_{a1}\vec{a_1}+n_{a2}\vec{a_2}))\right|^2 = \left|\frac{1-\exp(in_\infty\vec{q}\cdot\vec{a_1})}{1-\exp(i\vec{q}\cdot\vec{a_1})}\right|^2 \times \left|\frac{1-\exp(in_\infty\vec{q}\cdot\vec{a_2})}{1-\exp(i\vec{q}\cdot\vec{a_2})}\right|^2$$

$$= \left(\frac{\sin(n_\infty\vec{q}\cdot\vec{a_1})}{\sin(\vec{q}\cdot\vec{a_1})}\right)^2 \times \left(\frac{\sin(n_\infty\vec{q}\cdot\vec{a_2})}{\sin(\vec{q}\cdot\vec{a_2})}\right)^2$$

We have thus the well-known grating relation in optics which is non-zero if $\vec{q}\cdot\vec{a_1} = 2\pi[n]$ and $\vec{q}\cdot\vec{a_2} = 2\pi[n]$ imposing that $\vec{q}$ projected in the *(x,y)* plane is a vector of the reciprocal space. We call this vector or the reciprocal space G. Thus, the expression becomes:

$$S_{electron}(\vec{q},\theta_{\vec{q}}) = B\frac{1}{N}\left|\sum_{el_i=1}^{N_{el}\text{ in the primitive cell}} f_{el_i}(q) \sum_{i\in el_i} \exp(i[2\pi(hx_i+ky_i)+q_z z_i c])\right|^2$$

where $x_i$, $y_i$ and $z_i$ are the relative coordinates in the primitive cell. For graphene, it means 2 atoms in each plane and all planes stacked in the *z* direction.

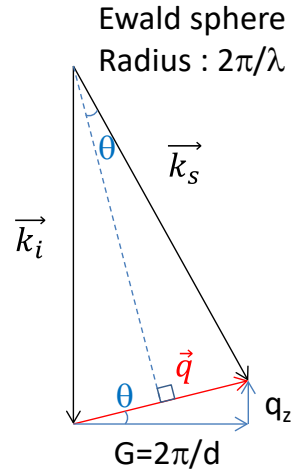

With simple algebra, we obtain:

$$q_z = \frac{2\pi}{\lambda} - \sqrt{\left(\frac{2\pi}{\lambda}\right)^2 - G^2}$$

We can introduce the atomic form factor for X-Rays (useful for conversion)

*Supplementary information*

$$f_{X-Ray}(q) = \sum_{i=1}^{4} a_i \, exp\left(-b_i \left(\frac{q}{4\pi}\right)^2\right) + c$$

…and for electrons:

$$f_e(q) = \sum_{i=1}^{5} a_i \, exp\left(-b_i \left(\frac{q}{4\pi}\right)^2\right)$$

with coefficients tabulated in the International Tables for Crystallography[2,3].

**IV-Graphene**

In the case of graphene, the distance in the plane ($l$ index is not including in our approach), we have:

$$d = \frac{1}{\sqrt{\frac{4}{3a^2}(h^2 - hk + k^2)}}$$

with $a = \sqrt{3} a_{cc} = 2.46 \text{Å}$

For the first ring, the indices are: 10, 11, 01, $\bar{1}0$, $\bar{1}\bar{1}$, $0\bar{1}$ corresponding to d = 2.126 Å

For the second ring, the indices are: 12, 21, $1\bar{1}$, $\bar{1}\bar{2}$, $\bar{2}\bar{1}$, $\bar{1}1$ corresponding to d = 1.227 Å

| d(Å) | 2.13 (ring 1) | 1.23 (ring 2) |
|---|---|---|
| q=2π/d (Å$^{-1}$) | 2.95 | 5.11 |
| f$_{X-Ray}$ | 3.11 | 1.92 |
| f$_{elec}$ | 1.25 | 0.59 |

**Table S1.** Parameters for graphene.

In order to show full DP, we have used XaNSoNS without any correction (i.e. for X-Ray) and report here the results for AA and AB at two wavelengths (0.173 Å and 0.037Å) corresponding to 0.5 keV and 100 keV (the limit values used in this study), respectively, considering the following formula:

*Supplementary information*

$$\lambda = \frac{h}{\sqrt{2meV}\sqrt{1+\frac{eV}{2mc^2}}}$$

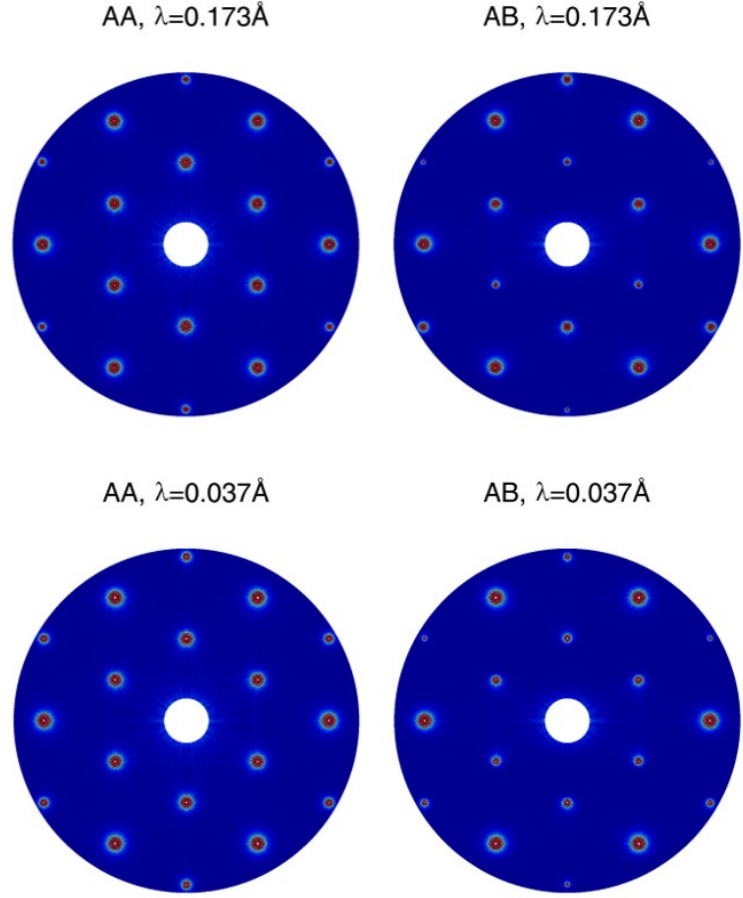

**Fig. S2**. DPs for two wavelengths (top: 5 keV; bottom: 100 keV) for AA (column 1) and AB (column 2) sequences.

Peak intensity ratios reported in **Table S2** below are obtained from XaNSoNs corrected with electronic atomic factor ($L_a = 30$ nm). Peaks are exactly at the right angular position and right $q$ position (checked by using a grid of $8192 \times 8192$) and have circular symmetry:

|     | 5kV | | 100kV | |
| --- | --- | --- | --- | --- |
|     | $I_{Ring2}/I_{Ring1}$ (with 3 weakest) | $I_{Ring2}/I_{Ring1}$ (with 3 strongest) | $I_{Ring2}/I_{Ring1}$ (with 3 weakest) | $I_{Ring2}/I_{Ring1}$ (with 3 strongest) |
| **A**   | 0;90 | 0.90 | 0.89 | 0.89 |
| **AA**  | 0.63 | 0.63 | 0.89 | 0.89 |
| **AB**  | 5.90 | 1.38 | 4.07 | 3.03 |
| **ABA** | 1.02 | 1.02 | 2.56 | 2.56 |
| **ABC** | 7.09 | 4.50 | 250  | 250  |

**Table S2.** Values extracted from XaNSoNS simulation.

If we do the same by using our model:

*Supplementary information*

|  | 5kV | | 100kV | |
|---|---|---|---|---|
|  | $I_{Ring2}/IR_{ing1}$ (with 3 weakest) | $I_{Ring2}/I_{Ring1}$ **(with 3 strongest)** | $I_{Ring2}/I_{Ring1}$ (with 3 weakest) | $I_{Ring2}/I_{Ring1}$ **(with 3 strongest)** |
| **A** | 0.89 | **0.89** | 0.89 | **0.89** |
| **AA** | 0.63 | **0.63** | 0.88 | **0.88** |
| **AB** | 5.99 | **1.37** | 4.10 | **3.04** |
| **ABA** | 1.02 | **1.02** | 2.57 | **2.57** |
| **ABC** | 7.24 | **4.5** | 333 | **333** |

**Table S3.** Values deduced from our simple model, to be compared to Table S2. The results are identical, proving the validity of our approach. Ratio values in bold are those reported in main Table1.

It is clear that the 6-fold symmetry is not valid for AB pair and a 3-fold symmetry shows up instead, using both calculation methodologies, i.e. XaNSoNS (**TableS2**) and our simpler approach (**TableS3**). Moreover, The Ring2/Ring1 intensity ratios compare fairly well, which validates our new methodology.

In this work, in addition to A and AA, we considered alternating AB and ABC sequences up to 6 layers by variation of 1 layer with our approach. What is interesting with it, is that any stacking sequence can be treated and easily obtained, for a large range of electron energy. Here are the additional plots of the peak intensity versus the electron energy which were not provided in the main text as **Fig.3**:

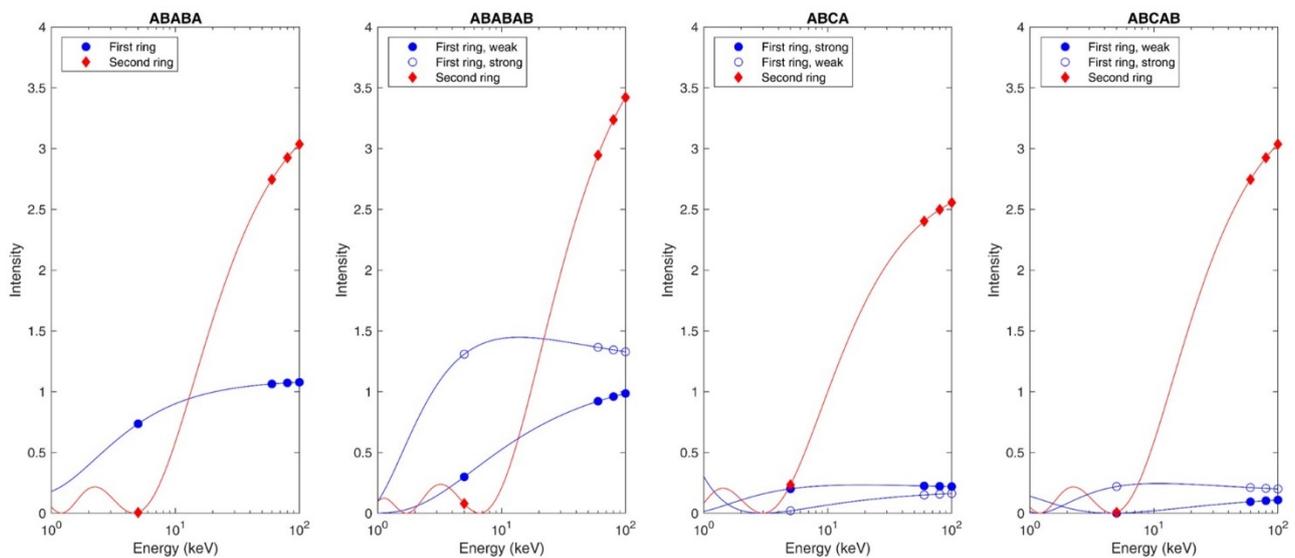

**Fig. S3.** Complementary data to main Fig.3. Intensity variations for a single spot from each ring, Ring1 (blue line) and Ring2 (red line), in calculated DPs with respect to the continuous variation of the electron energy (from 1 to 100 keV) for a selection of stacking configurations in graphene. When a three-fold symmetry of the intensity distribution appears on Ring1, two intensity values are provided corresponding to that of one weak spot (solid circle) and one strong spot (open circle), respectively. Electron energies corresponding to 5, 60, 80, and 100 kV TEM acceleration voltage are marked, for easier recognition.

Green-shaded boxes in **Table 1** of the main article indicate the combination of parameters which allows the various stacking sequences in graphene to be univocally discriminated. More in the details:

**A**: is the only case for which $I_{Ring1}$ and $I_{Ring2}$ are equal for both 5 and 100 keV; **AA**: is the only stacking sequence for which $I_{Ring2}$ is lower than $I_{Ring1(S)}$ at 5 keV while not showing any three-fold symmetry on Ring1. In case of doubt, the DP at 100 keV may be considered as a supplementary data, in order to check whether $I_{Ring1}$ and $I_{Ring2}$ are equal;

**AB**: exhibits a clear three-fold symmetry of spot intensity distribution ($I_{Ring1(S)}/I_{Ring1(W)} = 4.4$) and a slightly higher $I_{Ring2}$ over $I_{Ring1(S)}$ at 0.5 keV (1.4, although this intensity difference might not appear to eye), and a clearly higher $I_{Ring2}$ over $I_{Ring1(S)}$ at 100 keV ($I_{Ring2}/I_{Ring1(S)} = 3$);

**ABA**: is the only stacking sequence combining an absence of three-fold symmetry of the intensity distribution on Ring1 while $I_{Ring1}$ and $I_{Ring2}$ are equal at 5 keV, and a clearly higher $I_{Ring2}$ over $I_{Ring1}$ at 100 keV ($I_{Ring2}/I_{Ring1} = \sim 3$);

**ABAB**: distinguishes from the other sequences of the series the same way as AB does, but can be discriminated from the latter by considering $I_{Ring2}/I_{Ring1(S)}$ whose value is reversed (0.2 instead of 1.4) with $I_{Ring2}$ as low as $I_{Ring1(W)}$;

**ABABA**: is the first stacking sequence for which Ring2 vanishes at 5 keV, and the only one which does not exhibit a three-fold symmetry of the intensity distribution on Ring1 at 100 keV meanwhile;

**ABABAB**: is the only stacking sequence showing a three-fold symmetry of spot intensity distribution on Ring1 ($I_{Ring1(S)}/I_{Ring1(W)} = 4.4$) while Ring2 almost vanishes, at 5 keV; **ABC**: is the only configuration exhibiting a strong intensity difference between Ring1 and Ring2 ($I_{Ring2}/I_{Ring1} = 4.5$) at 5 keV, while Ring1 vanishes at 100 keV;

**ABCA**: is the only stacking sequence showing an obvious three-fold symmetry of the intensity distribution on Ring1 ($I_{Ring1(S)}/I_{Ring1(W)} = 10$, while half of the spots are so weak that they might barely appear) at 5 keV, while showing a strong intensity difference between Ring2 and Ring1 at 100 keV ($I_{Ring2}/I_{Ring1} = \sim 12$);

**ABCAB**: is characterised by a combination of a vanishing of half of the spots on Ring1, inducing a strong three-fold symmetry of the intensity distribution, and a vanishing of Ring2 at 5 keV. In case of doubt, the DP at 100 keV may be considered as a supplementary data, in order to check whether $I_{Ring2}$ is strongly higher than $I_{Ring1}$ ($I_{Ring2}/I_{Ring1} = 15$);

**ABCABC**: shows the lowest $I_{Ring2}/I_{Ring1(S)}$ ratio value (0.4) of the series with a very weak Ring2 at 0.5 keV, combined with the reverse situation at 100 keV, i.e., $I_{Ring2}/I_{Ring1} = 285$ with Ring1 being barely visible.

**V- MoS$_2$**

In the case of MoS$_2$, the distance in the plane (*l* index is no longer valid in our approach), we have:

$$d = \frac{1}{\sqrt{\frac{4}{3a^2}(h^2 - hk + k^2)}}$$

with $a = \sqrt{3}a_{cc} = 3.22$Å

For the first ring, the indices are: 10, 11, 01, $\bar{1}0$, $\bar{1}\bar{1}$, $0\bar{1}$ corresponding to d = 2.789 Å

For the second ring, the indices are: 12, 21, $1\bar{1}$, $\bar{1}\bar{2}$, $\bar{2}\bar{1}$, $\bar{1}1$ corresponding to d = 1.610 Å

| d(Å) | 2.79 (first ring) | 1.610 (second ring) |
|---|---|---|
| q=2π/d (Å$^{-1}$) | 2.25 | 3.90 |
| f$_{elec}$ (Mo;S), real part | 5.74;3.21 | 3.49;1.77 |

**Table S4.** Parameters for MoS$_2$.

In order to take into account the imaginary part of the complex atomic factor $f_e = f_{e,r} + if_{e,i}$, we have assumed similar behaviours for X-Rays and electrons and extended to 100keV the law reported by Henke *et al*[4] after the absorption edges as illustrated by **Fig.S4**, giving the following corrections:

$f_{e,S} = f_{e,S,real} \times \{1 + i/16 \times 0.42 \times [E/2.5]^{-1.7}\}$ if E < 2.5 kV

$f_{e,S} = f_{e,S,real} \times \{1 + i/16 \times 4.22 \times [E/2.5]^{-1.67}\}$ if E > 2.5 kV

and

$f_{e,Mo} = f_{e,Mo,real} \times \{1 + i/42 \times 3.10 \times [E/2.6]^{-1.36}\}$ if E < 2.6 kV

$f_{e,Mo} = f_{e,Mo,real} \times \{1 + i/42 \times 17.36 \times [E/2.6]^{-1.36}\}$ if 20.4 > E > 2.6 kV

$f_{e,Mo} = f_{e,Mo,real} \times \{1 + i/42 \times 3.5 \times [E/20.4]^{-1.6}\}$ if E > 20.4 kV (estimated)

*Supplementary information*

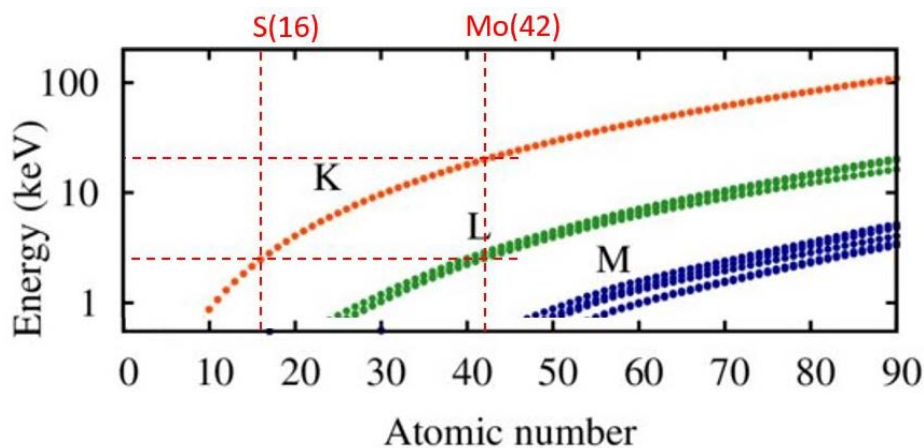

**Fig. S4.** Plots of the energy values for the various adsorption edges K, L, M in function of the atomic number. For sulfur and molybdenum (indicated along with their atomic number) as in $MoS_2$, only L and K adsorption edges have to be considered. For $WS_2$ for instance (Z of W = 74), M adsorption edge should additionally be considered. Modified from Robinson[5]